\documentclass[onecolumn,pra,aps,epsf,amsmath,amssymb,showpacs,preprint]{revtex4-1}
\usepackage{comment}
\usepackage[dvipsnames]{xcolor}

\usepackage{yfonts}

\usepackage{graphicx}
\usepackage{xcolor}
\usepackage{dcolumn}
\usepackage{bm}

\def\bea{\begin{equation}\begin{aligned}}
\def\eea{\end{aligned}\end{equation}}

\newcommand{\bfred}[1]{\color{red} #1}

\begin{document}
\title{Macroscopic Ferromagnetic Dynamics}
\author{Chen Sun}
\email{chensun@hnu.edu.cn}
\affiliation{ School of Physics and Electronics, Hunan University, Changsha 410082, China. }
\author{Wayne M. Saslow}
\email{wsaslow@tamu.edu}
\affiliation{ Texas A\&M University, College Station, Texas, 77843, U.S.A. }

\begin{abstract}
In metals with finite magnetization $\vec{M}$, experiment shows that transverse polarized dc spin currents $\vec{J}_{i}$ both decay and precess on crossing a finite sample thickness. The present work uses Onsager's irreversible thermodynamics, with $\vec{M}$ and $\vec{J}_{i}$ as fundamental variables, to develop a theory with aspects of the Landau-Lifshitz theory solely for the $\vec{M}$ of (charged) electronic ferromagnets, and of the Leggett theory for the $\vec{M}$ and $\vec{J}_{i}$ of (uncharged) nuclear paramagnets. As for the ferromagnet of Landau-Lifshitz, $\partial_{t}\vec{M}$ includes a characteristic decay time $\tau_{M}$. As for the nuclear paramagnet, $\partial_{t}\vec{J}_{i}$ includes a characteristic decay time $\tau_{J}$, is driven by the gradient of a (vector) spin pressure, and precesses about a mean-field proportional to $\vec{M}$. The spin pressure has a coefficient $G$ proportional to a velocity squared, and $D_{0}\equiv \frac{1}{2}G\tau_{J}$ serves as an effective diffusion coefficient.  These equations apply when spin currents are generated. Using the derived dynamical equations for the magnetization and for the spin current, we obtain the steady state (dc limit) solution whose transverse wavevector squared is complex, with real part from diffusion and imaginary part from precession.  The ac case is also considered.
\end{abstract}




\date{\today}

\maketitle

\section{Introduction}
\label{s:Intro}
The field of spintronics uses spin currents to transfer spin (properly, magnetization $\vec{M}$) from one material to another, permitting devices to both read and write magnetic information.
On the one hand, ferromagnetic resonance might drive spin currents $\vec{J}_{i}$ (properly, magnetization currents) out of a ferromagnet (spin pumping); on the other hand, spin currents might enter and transfer their spin to a ferromagnet (spin transfer torque). However, spin currents are not included in the classic work of Landau and Lifshitz,\cite{LLMagnetics35} so there is a need to develop the spin current dynamics.

A theory of spin transport in ferromagnets was developed earlier by Zhang, Levy and others \cite{ZhangLevyFert02,ZLZA04} using a phenomenological two-magnetization ($\vec{M}$, $\vec{m}$) model for the spin dynamics.  For this model, $\vec{M}$ is the local equilibrium magnetization and $\vec{m}$ is the out-of-equilibrium magnetization. For transverse spin diffusion this model finds a complex decay length and wavevector involving exchange-field precession, magnetization decay, and spin current diffusion. Specifically, Ref.~\cite{ZhangLevyFert02} obtains the square of the wavevector for transverse spin waves
\begin{align}
k^2=-\frac{1}{2D_0\tau_{sf}} \pm i\frac{J}{2  h D_0};
\label{eq:ZLF}
\end{align}
here $\tau_{sf}$ is the spin-flip relaxation time, $D_0$ the spin diffusion constant, and $J$ the interaction strength between the two magnetizations (proportional to the usual magnetization $M$).\cite{{twomagvariables}}  No external magnetic field was considered.  This ($\vec{M}$, $\vec{m}$) model has been applied successfully by Taniguchi and others\cite{TaniguchiAPEx08} to explain spin transport phenomena observed experimentally.

We have previously applied irreversible thermodynamics based on the ($\vec{M}$, $\vec{m}$) variables.\cite{Saslow17}  However, Leggett's work on paramagnetic liquid 3He at low temperatures\cite{Leggett70} shows that the variables ($\vec{M}$, $\vec{J}_{i}$) have a more natural interpretation in terms of Fermi liquid theory.

This has motivated the present work, which for a ferromagnet applies Onsager's irreversible thermodynamics to the variables ($\vec{M}$, $\vec{J}_{i}$); it also applies the theory to study the precession and decay of such spin currents. A separate work extends Leggett's application of Fermi liquid theory for a (nuclear) paramagnet\cite{Leggett70} to the case of a two-band (up-down) ferromagnet.\cite{SasSunFLFM}  The present theory and the Fermi liquid theory for the ferromagnet have equations of motion with the same structure, although for Fermi liquid theory the coefficients are more specified. The theory applies to both conductors and insulators; for conductors we neglect charge currents. Because insulators are subject to heating from spin currents, but not from charge currents, there is considerable interest in spin currents in insulators, to which the Onsager theory applies but not Fermi liquid theory.\cite{Brataas20}

\begin{figure}[!htb]
  \scalebox{0.4}[0.4]{\includegraphics{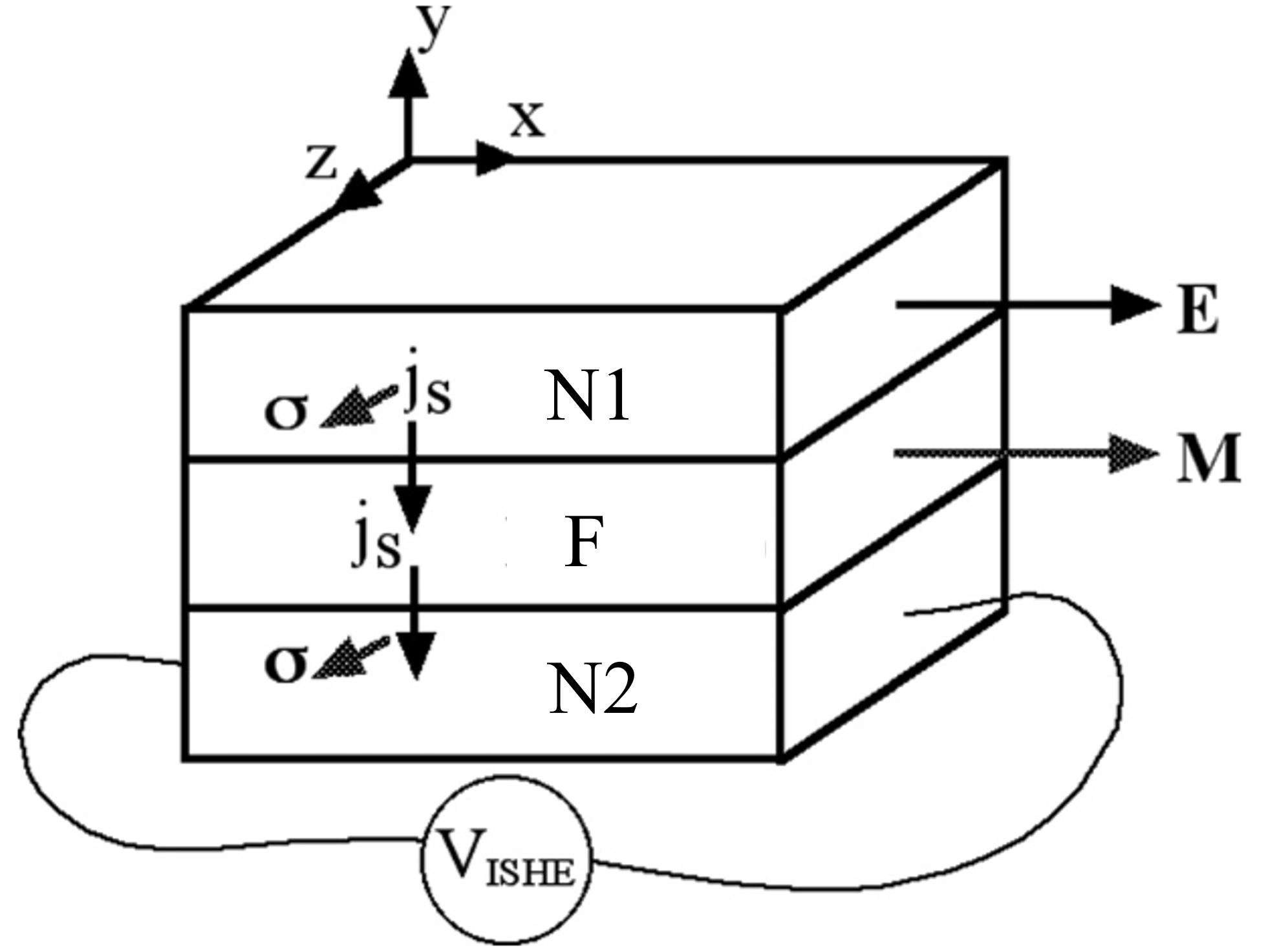}}
\caption{  An experimental geometry generating a spin current from a normal material N1 to a ferromagnet F (with magnetization $\vec{M}$ along $\hat{x}$) whose thickness can be varied, and then to the normal material N2 where the inverse spin Hall effect produces a measured voltage.  The spin current bi-vector has, in this special case, real-space flow $j_{s}$  along $-\hat{y}$ and spin polarization $\vec{\sigma}$ along $\hat{z}$, and thus is polarized transverse to $\vec{M}$.  A voltage across the upper layer produces $\vec{E}$ along $\hat{x}$ that, by the spin Hall effect, produces $j_{s}$ with $\vec{\sigma}$ that is normal to the magnetization $\vec{M}$ along $\hat{x}$ that enters the ferromagnet.  The spin current crosses F1 and then enters N2, where it generates the voltage $V_{ISHE}$ by the inverse spin Hall effect.}
\label{fig:geometry}
\end{figure}

Fig.~\eqref{fig:geometry} gives a relevant experimental geometry for the study of precession and decay of transverse spin currents. It first uses the spin Hall effect in one material to produce spin current from a charge current, and then uses the inverse spin Hall effect in a second material to produce a (measurable) charge current from the spin current.


Alternatively, for the same geometry a temperature gradient along $\hat{y}$, would drive a spin current along $\hat{y}$ with its magnetization along $\hat{z}$.\cite{Chien13}


Using the equations of motion derived in Sect.~\ref{s:eqsmo}, Sect.~\ref{s:statics} derives Eq.~\eqref{kT2}, which is very similar to Eq.~\eqref{eq:ZLF}.  Despite this agreement, the ($\vec{M}$, $\vec{m}$) theory of Ref.~\cite{ZhangLevyFert02} and ($\vec{M}$, $\vec{J}_{i}$) of the present work are not equivalent.  This is in part because, as in the Fermi liquid theory, the present theory has a vector pressure-gradient-like term in $\partial_{t}\vec{J}_{i}$ that does not appear in the ($\vec{M}$, $\vec{m}$) theory. In addition, spin diffusion does not appear explicitly in the $\vec{M}$, $\vec{m}$ theory; the apparent diffusion is a combination of the decay time for $\vec{M}$ and the coefficient $G$ of the gradient of the pressure term for $\vec{J}_{i}$, where $G$ is proportional to the Fermi velocity squared.  The vector ``spin pressure'' is present even without Fermi liquid interactions.

\subsection{Methods}
\label{ss:Methods}
Many authors have previously applied Onsager's irreversible thermodynamics to develop a theory for ferromagnets using only $\vec{M}$, and thus without spin currents, and obtained the Landau-Lifshitz equation, not the Landau-Lifshitz-Gilbert equation.\cite{SasRivkin08} The present work can be thought of as an extension of the irreversible thermodynamics of Johnson and Silsbee\cite{JohnsonSilsbee87} to include the dynamics of the magnetization current $\vec{J}_{i}$, where the index $i$ gives the direction of flow and the vector part gives the direction in which the magnetization  points.

The resulting equations of motion for $\vec{M}$ and $\vec{J}_{i}$ have the same symmetry as in the equations in Leggett's Fermi liquid theory, but
also include decay of $\vec{M}$ with a characteristic time $\tau_{M}$ that corresponds to $\tau_{sf}$ of \eqref{eq:ZLF}.\cite{LeggettMdecay} It also replaces Fermi liquid coefficients by other dimensionless and more phenomenological parameters.  Note that for small deviations of $\vec{M}$ from local equilibrium $\vec{M}_{le}$, the form $\partial_{i}(\vec{M}-\vec{M}_{le})$ appears in the Fermi liquid theory for $\partial_{t}\vec{J}_{i}$.\cite{BaymPethick78}  This requires relaxation of $\vec{M}$, either by decay (which for liquid $^{3}$He at low temperatures can be hours or even days) or by the much slower process of diffusion.  By simply permitting $\vec{M}$ to decay, the linearized equations give a dc wavevector and decay length $l$ similar to that of \eqref{eq:ZLF}, corresponding to oscillation and decay from surfaces.

Recall that Onsager's irreversible thermodynamics can be expected to apply only near equilibrium and at frequencies low relative to a characteristic inverse collision time.  Implicit to the Onsager theory is a third, and relatively short, local equilibrium decay time $\tau_{le}$ that is relevant to attaining local equilibrium before equilibrating either $\vec{M}$ or $\vec{J}_{i}$, so we assume that $\tau_{le}\ll\tau_{J},\tau_{M}$.
Such a separation of time scales between $\tau_{le}$ and other characteristic times is considered a necessary condition for the existence of a local-equilibrium distribution \cite{Zubarev}.
For ferromagnets the condition $\tau_{le}\ll \tau_{M}$ is satisfied whenever spin-orbit coupling is weak, since the decay of $\vec{M}$ is governed by the spin-orbit coupling. However, we do not know a simple argument to  rigorously establish that $\tau_{le}\ll \tau_{J}$; in reality whether this condition is satisfied may depend on the materials. The derived equations of motion should apply for frequencies $\omega$ subject to $\omega\tau_{le}\ll1$.

Related studies of transverse spin wave modes for paramagnetic conductors, based on Silin's implementation of Landau's Fermi liquid theory, \cite{LandauFLT56,LandauFLT57,Silin1,Silin2,Silin3} have been made by Platzman and Wolff,\cite{PlatzmanWolff} and by Fredkin and Wilson.\cite{WIlsonFredkin70}  These works, for paramagnets, and the related work of Leggett, do not consider decay of $\vec{M}$.  Such decay, however, is present in the theory of Landau and Lifshitz for ferromagnets.\cite{LLMagnetics35}  Recent theoretical studies of spin currents in ferromagnets include phenomenological approaches to magnetic damping and spin diffusion \cite{Tserkovnyak09} and the proposal that different forms of the spin-orbit interaction lead to new forms for the spin currents.\cite{TGS15,ALSH19}  Ref.~\cite{Kim24} provides a recent review.
\\

This paper is organized as follows. Sect.~\ref{s:thermovar} presents the thermodynamics of the system. Sect.~\ref{s:eqsmo} gives the equations of motion but not the decay terms or fluxes.  Sect.~\ref{s:fluxforce} applies Onsager's irreversible thermodynamics to obtain the decay terms and fluxes.  Sect.~\ref{s:statics} solves the equations of motion in the static limit. Sect.~\ref{s:finitefrequencyLT} studies the finite frequency longitudinal and transverse modes.
Sect.~\ref{s:FLcoefs}  studies the finite frequency longitudinal spin waves for paramagnets using known values for the Fermi liquid coefficients for some paramagnetic alkali metals.  Sect.~\ref{s:summary} provides a summary and conclusions.

\section{Variables, Thermodynamics, Symmetries}
\label{s:thermovar}
For specificity we take gyromagnetic ratio $-\gamma$ with $\gamma>0$, as for electrons.  Reversing the sign of $\gamma$ yields the case of liquid $^{3}$He.  We ignore effective fields due to anisotropy and exchange, but they can easily be incorporated. Our focus is mainly on ferromagnets and spin currents that do not depend on spin-orbit effects.  For simplicity we do not include spin flow or heat flux due to inhomogeneous magnetization or inhomogeneous temperature. For paramagnets, these effects are possible due to spin-orbit interaction and spin-dependent scattering.\cite{Dyakonov2017} Strictly speaking, at least some spin-orbit interaction is necessary for the magnetization to decay, but we assume the spin-orbit interaction is weak enough to ignore its contributions to spin currents (see, for example, the spin currents in \cite{TGS15,ALSH19}).

\subsection{Variables and Thermodynamics}
\label{ss:Variables}
The magnetic energy density for the interaction of a ferromagnet of magnetization $\vec{M}$ with an external field $\vec{B}_{e}$ is
\begin{equation}
\varepsilon_{ext}=-\vec{M}\cdot\vec{B}_{e},
\label{varext}
\end{equation}
and for $M$ not far from $M_{0}$ we model the effective internal energy density as
\begin{equation}
\varepsilon\approx\frac{(M-M_{0})^{2}}{2\chi}\equiv \frac{ (\delta M)^{2}}{2\chi}.
\label{vareps}
\end{equation}

At fixed $\vec{M}_{0}$, we define the internal magnetic field $\vec{B}$  in terms of the magnetization $\vec{M}$ by
\begin{equation}
\vec{M}=\vec{M}_{0}+\chi\vec{B}, \qquad \delta\vec{M}=\chi {  \vec{B}}.
\label{vecM}
\end{equation}
For collinear vectors, we have
\begin{equation}
B\equiv\frac{\partial\varepsilon}{\partial M}\approx \frac{M-M_{0}}{\chi}.
\label{B}
\end{equation}
In equilibrium we minimize $\varepsilon_{ext}+\varepsilon$ with respect to $M$ to obtain $B=B_{e}$, so in equilibrium $M=M_{0}+\chi B_{e}$.  Note that $M$ has units of $\gamma\hbar n$, where $n$ is a number density. Spin density has units of $\hbar n$.

We now make a formal analogy between the magnetization $\vec{M}$ and the external field $\vec{B}_{e}$ (vectors in spin space), and the spin density flux $\vec{J}_{i}$ and the (fictitious) external spin current field $\vec{\Psi}_{e,i}$ (bi-vectors of spin order and flow in real space).\cite{fictitious}  Our goal is to provide a macroscopic description of the decay in space of the variables $\vec{M}$ and $\vec{J}_{i}$, as in eq.(1).

Because $\vec{J}_{i}$ is a magnetization flux, it has units of $(\gamma\hbar n)v$, where $v$ is a velocity.  We now assume an internal energy density $\varepsilon'=(\beta/2)|\vec{J}_{i}|^{2}$, where with $m$ the carrier mass, $\beta$
 has units of $mnv^{2}/(\gamma\hbar nv)^{2}$, or $m/[n(\gamma\hbar)^{2}]$.
We also take an interaction energy density $\varepsilon'_{ext}=-\vec{J}_{i}\cdot\vec{\Psi}_{ei}$, where the (imaginary) symmetry-breaking external field $\vec{\Psi}_{e,i}$ has units of $(m/\gamma\hbar)v$.  Thus we can define the more physically interpretable $\vec{v}_{e,i}\equiv(\gamma\hbar/2m)\vec{\Psi}_{e,i}$, and correspondingly $\vec{v}_{i}\equiv(\gamma\hbar/2m)\vec{\Psi}_{i}$.  In local equilibrium, $\vec{\Psi}=\vec{\Psi}_{e,i}$.

Minimizing $\varepsilon'_{ext}+\varepsilon'$ with respect to $\vec{J}_{i}$ gives the local equilibrium state property $\vec{J}^{(le)}_{i}=(1/\beta)\vec{\Psi}_{e,i},$
so we introduce 
the quantity $\rho_{M}$ by defining
\begin{equation}
\vec{J}_{i}=\frac{1}{\beta}\vec{\Psi}_{i}=\frac{1}{\beta}\frac{2m}{\gamma\hbar}\vec{v}_{i}\equiv\rho_{M}\vec{v}_{i}, \qquad \rho_{M}\equiv \frac{1}{\beta}\frac{2m}{\gamma\hbar}.
\label{Ji}
\end{equation}
From the units of $\beta$, $\rho_{M}$ has units of $\gamma\hbar n$, or a magnetization.

We now introduce $\rho_{m}$ by requiring that
\begin{equation}
\frac{\rho_{m}}{2}|\vec{v}_{i}|^{2}\equiv\frac{\beta}{2}|\vec{J}_{i}|^{2}=\frac{\beta}{2}\rho_{M}^{2}|\vec{v}_{i}|^{2},
\label{rhomM}
\end{equation}
where the second equality arises from \eqref{Ji}. Thus
\begin{equation}
\rho_{m}=\beta\rho_{M}^{2}=\frac{2m}{\gamma\hbar}^{2}\frac{1}{\beta}=\frac{2m}{\gamma\hbar}\rho_{M}. 
\label{rhomM2}
\end{equation}
From the units of $\beta$, $\rho_{m}$ has units of $mn$, or a mass density. 

Consistent with the actual magnetic field $\vec{B}_{e}$ energy and the internal energy, but with no imaginary field $\vec{\Psi}_{ei}$ energy, for the total thermodynamic energy density differential $du$ we take
\begin{equation}
  du=d\varepsilon_{ext}+d\varepsilon=Tds+  (\vec{B}-\vec{B}_{e}) \cdot d\vec{M} +\frac{2m}{\gamma\hbar}\vec{v}_{i}\cdot d\vec{J}_{i}.
\label{deps}
\end{equation}

For small deviations from equilibrium of only the magnetic variables (so $ds=0$), the energy change is
\begin{equation}
\Delta\varepsilon=\frac{1}{2\chi}(\Delta\vec{M})^{2}+\frac{1}{2\rho_{M}}
(\frac{2m}{\gamma\hbar}) (\vec{J}_{i})^{2}.
\label{eps}
\end{equation}
Both $\chi$ and $\rho_{M}=\rho_{m}(\gamma\hbar/2m)$ are response functions for which simple expressions in terms of thermal fluctuations are derived in the Appendix.

\subsection{Symmetries}
\label{ss:Symmetries}
In the absence of spin-orbit effects, so the internal energy $\varepsilon$ is invariant under virtual rotations of all spin-space vectors by $\delta\vec{\theta}$, as in magnetization change $\delta\vec{M}=\delta\vec{\theta}\times\vec{M}$, we have
\begin{equation}
  \vec{B} \times\vec{M}+\frac{2m}{\gamma\hbar}\vec{v}_{i}\times\vec{J}_{i}=\vec{0}.
\label{depsrot}
\end{equation}
This implies
\begin{equation}
\vec{v}_{i}\cdot(\vec{J}_{i}\times\vec{B})=0, \quad \vec{v}_{i}\cdot(\vec{J}_{i}\times\vec{M})=0,
\label{constraint}
\end{equation}
results that will simplify some of the calculations to follow.  Additional relations follow from \eqref{depsrot} but they will not be needed.

In the absence of dissipation, under time-reversal $\vec{B} $ and $\vec{M}$ are odd, and $\vec{v}_{i}$ and $\vec{J}_{i}$ are even.

\section{Equations of Motion}
\label{s:eqsmo}
Onsager's irreversible thermodynamics involves unknown fluxes (we typically employ upper case $J$'s to avoid confusion with the lower case space index $j$) and unknown sources (generically $R$'s).
However, following Ref.~\cite{BaymPethick78}, instead of the ``flux of the flux density'' written as $\vec{J}_{ij}$ we use $\vec{\Pi}_{ij}$ -- a spin-space stress tensor.  The general approach enforces the condition that
\eqref{deps} remains true at all times, and that the rate of entropy production is always non-negative.  Note that the dimensions of $\vec{J}_i$ and $\vec{\Pi}_{ij}$ are $[\vec{J}_i]=[M v]$ and $[\vec{\Pi}_{ij}]=[\vec{J}_i v]=|Mv^{2}|$, respectively.

In terms of these quantities we take the equations of motion (the first of which is also called a continuity equation) to have the form
\begin{eqnarray}
\partial_{t}u&+&\partial_{i}J^{u}_{i}=0.\label{depsdt}\\
\partial_{t}s&+&\partial_{i}J^{s}_{i}=R^{s}\ge0.\label{dsdt}\\
\partial_{t}\vec{M}&+&\partial_{i}\vec{J}_{i}=-\gamma\vec{M}\times\vec{B}_{e}+\vec{R}^{M}
.\label{dMdt}\\
\partial_{t}\vec{J}_{i}&+&\partial_{j}\vec{\Pi}_{ij}=-\gamma\vec{J}_{i}\times\vec{B}_{e}-\gamma\vec{J}_{i}\times \lambda\vec{M}+\vec{R}^{\vec{J}}_{i}
.\quad\label{dJMdt}
\end{eqnarray}
Both $\vec{M}$ and $\vec{J}_{i}$ precess around the external field $\vec{B}_{e}$, and we include the symmetry-allowed precession of $\vec{J}_{i}$ about a mean-field we write as $\lambda\vec{M}$, where $\lambda$ can be computed using Fermi liquid theory. \cite{Leggett70,BaymPethick78} Note that the precession times associated with $\vec B_{e}$ and $\vec B_{e} +\lambda\vec{M}$ must also be much larger than $\tau_{le}$ for Onsager's irreversible thermodynamics to apply. This is satisfied when  $\vec B_{e} +\lambda\vec{M}$ is not large. The fact that the spin echo experiments in the nuclear paramagnet $^{3}$He were successfully interpreted using Leggett's theory
indicates the correctness of this assumption for that system. We make the same assumption in the present case.
$\partial_{t}\vec{M}$ can contain an effective field term in $\vec{B}$ due to non-uniform exchange, with spatial derivatives of the magnetization, and causing spin wave dispersion.  For simplicity this term has been omitted.\cite{LLMagnetics35}

We now perform the crucial step of using the equations of motion and the thermodynamics to derive the rate of heat production $TR^{s}$ (see e.g. \cite{Saslow17} for more details of such a derivation).  We then enforce the condition that $TR^{s}$ is always non-negative.
On rewriting \eqref{deps} using all of the above equations of motion, $TR^{s}$ satisfies
\begin{eqnarray}
0\le TR^{s}&=&T\partial_{i}J^{s}_{i}+T\partial_{t}s\cr
&=&\partial_{i}(TJ^{s}_{i})-J^{s}_{i}\partial_{i}T+
{ \partial_{t}u - (\vec{B}-\vec{B}_{e})\cdot\partial_{t}\vec{M} }
-\frac{2m}{\gamma\hbar}\vec{v}_{i}\cdot\partial_{t}\vec{J}_{i}.
\label{1st}
\end{eqnarray}

The time-derivatives above are given by \eqref{dsdt}, \eqref{dMdt}, and \eqref{dJMdt}. Because of \eqref{depsrot} the contributions to $TR^{s}$ of two precession terms (the precession term in $\partial_{t}\vec{M}$ and the first precession term in $\partial_{t}\vec{J}_{i}$) cancel, and by the second part of \eqref{constraint} the contributions to $TR^{s}$ of the second precession term in $\partial_{t}\vec{J}_{i}$ is zero. Thus the precession terms do not contribute to $TR^{s}$.

Collecting the remaining terms in $TR^{s}$ yields a sum of terms in the form of \\
(a) a divergence involving fluxes, \\
(b) a sum of products of fluxes and their thermodynamic forces, and \\
(c) a sum of products of sources and their thermodynamic forces.  \\

In \eqref{deps} the term in $\vec{B}^{*}\cdot d\vec{M}$ has sign opposite to the other terms, a feature that appears in the divergence terms, the flux-related entropy production terms, and the source-related entropy production terms.

\subsection{Divergence terms}
\label{ss:divterms}
In $TR^{s}$ the flux terms sum to
\begin{equation}
-\partial_{i}[J^{u}_{i}-TJ^{s}_{i}
-(\vec{B}-\vec{B}_{e})\cdot\vec{J}_{i}-\frac{2m}{\gamma\hbar}\vec{v}_{j}\cdot\vec{\Pi}_{ji}].
\label{fluxes}
\end{equation}
This will be set to zero, thus determining $J^{u}_{i}$ in terms of the other fluxes. (An arbitrary curl, whose divergence is thus zero, is still permitted in $J^{u}_{i}$, but we know of no physics associated with it.  We will not actually need $J^{u}_{i}$.) These fluxes, in the linear response regime,  are proportional to the thermodynamic forces with coefficients perhaps associated with the order parameter $\vec{M}$.

\subsection{Product of diffusion terms}
\label{ss:proddiffterms}
In $TR^{s}$ the terms due to products of fluxes and thermodynamic forces sums to
\begin{equation}
-J^{s}_{i}\partial_{i}T
  -\vec{J}_{i}\cdot\partial_{i}(\vec{B}-\vec{B}_{e}) -\vec{\Pi}_{ji}\cdot \frac{2m}{\gamma\hbar}\partial_{i}\vec{v}_{j}\ge0.
\label{flux-force}
\end{equation}
We will take the fluxes to be proportional to (as consistent with symmetry) a sum of thermodynamic forces.

If $\partial_{i}(\vec{B}-\vec{B}_{e})\ne0$, then the system has too much or too little magnetization and, if it cannot adjust by magnetization decay to or from the system, then it will adjust by magnetization diffusion.


\subsection{Reversible flux terms}
\label{ss:Revfluxterms}
The intrinsic signature of $\vec{J}_{i}$ under time-reversal is even because it is a flow (involving a velocity) of $\vec{M}$.  From our considerations on $\vec{J}_{i}$ we can give it an irreversible (dissipative) part by writing
\begin{equation}
\vec{J}_{i}=\rho_{M}\vec{v}_{i}+\vec{J}^{\,'}_{i},\qquad \rho_{M}\equiv\rho_{m}(\frac{\gamma\hbar}{2m}),
\label{eq:J}
\end{equation}
where $\vec{J}^{\,'}_{i}$ is dissipative.

Use of the reversible term of \eqref{eq:J} in \eqref{flux-force} indicates there can be, with a presently  unspecified coefficient $\alpha$,
a reversible term in $\vec{\Pi}_{ij}$ that is proportional to $(\vec{B}-\vec{B}_{e})\delta_{ij}$.
With the remaining parts of $\vec{\Pi}_{ji}$ and $\vec{J}_{i}$ given as $\vec{\Pi}^{'}_{ji}$ and $\vec{J}^{\,'}_{i}$, we write
\begin{equation}
\vec{\Pi}_{ji}=\vec{\Pi}^{'}_{ji} -\alpha(\vec{B}-\vec{B}_{e})\delta_{ij}.
\label{Pij-J}
\end{equation}
The reversible terms cannot contribute to the rate of entropy production; as a consequence, they are related.

We now use \eqref{Pij-J} in \eqref{flux-force} to replace $\vec{\Pi}_{ji}$ by $\vec{\Pi}^{\,'}_{ji}$ and $\vec{J}_{i}$ by $\vec{J}^{\,'}_{i}$.  This gives \eqref{flux-force} the terms
\begin{eqnarray}
&&-\rho_{M}\vec{v}_{i}\cdot\partial_{i}(\vec{B}-\vec{B}_{e})+\frac{2m}{\gamma\hbar}\alpha(\vec{B}-\vec{B}_{e})\cdot\partial_{i}\vec{v}_{i}\cr
&&=-\partial_{i}[\rho_{M}\vec{v}_{i}\cdot(\vec{B}-\vec{B}_{e})]+\left(\rho_{M}+\frac{2m}{\gamma\hbar}\alpha\right)(\vec{B}-\vec{B}_{e})\cdot\partial_{i}\vec{v}_{i}.\qquad
\label{extra}
\end{eqnarray}
In the above equation the first term is a divergence, and must be added to \eqref{fluxes}; this does not affect the equations of motion for $\vec{M}$ and $\vec{J}_{i}$, so we do not consider it further. The last two terms are proportional to $(\vec{B}-\vec{B}_{e})\cdot\partial_{i}\vec{v}_{i}$, which is odd under time-reversal.  They cannot contribute to the rate of entropy production, which is even under time-reversal.  Therefore they must sum to zero, to give the constraint
\begin{equation}
\alpha=-\frac{\gamma\hbar}{2m}\rho_{M}.
\label{rho-alpha}
\end{equation}

\subsection{Product of decay terms}
\label{ss:Prodecayterms}
If $\vec{B}-\vec{B}_{e}\ne0$, then the system has too much or too little magnetization, and can adjust by magnetization decay to or from the system.

The entropy production terms due to products of sources and thermodynamic forces sum, on using $\vec{B}\cdot(\vec{M}\times\vec{B})=0$ and \eqref{constraint}, to
\begin{equation}
 -(\vec{B}-\vec{B}_{e})\cdot\vec{R}^{M}
-\frac{2m}{\gamma\hbar}\vec{v}_{i}\cdot\vec{R}^{J}_{i}\ge0,
\label{source-forceD}
\end{equation}
where, by the Onsager principle for off-diagonal dissipative terms, $\vec{R}^{M}$ and $\vec{R}^{J}_{i}$ must be chosen to make the two terms equal and non-negative.

\section{Thermodynamic Fluxes and Forces}
\label{s:fluxforce}
For this system the thermodynamic forces are $\partial_{i}T$, $\vec{B}-\vec{B}_{e}$, $\partial_{i}(\vec{B}-\vec{B}_{e})$,\cite{Torrey56} $\vec{v}_{j}$ and $\partial_{i}\vec{v}_{j}$.
In \eqref{flux-force}, on retaining only the diagonal terms, for the dissipative, or irreversible, thermodynamic fluxes (expressed in terms of thermodynamic forces) we have
\begin{eqnarray}
J^{s}_{i}&=&-\frac{\kappa}{T}\partial_{i}T, \label{Jsi}\\
\vec{J}^{\,'}_{i}&=&-D_{M}\chi\partial_{i}(\vec{B}-\vec{B}_{e}), \label{vecJi}\\
\vec{\Pi}^{'}_{ji}&=&-D_{\vec{J}}\rho_{M}\partial_{i}\vec{v}_{j}. \label{vJji}
\end{eqnarray}
In the above, both $\chi$ and $\rho_{M}$ are non-negative equilibrium parameters, and
$D_{M}$ and $D_{\vec{J}}$ are new diffusion constants associated with $\vec{M}$ and $\vec{J}_{i}$.  To ensure non-negative entropy production, the parameters $D_{M}$, $D_{\vec{J}}$ and the thermal conductivity $\kappa$ must be non-negative.

In \eqref{source-forceD}, on retaining only the diagonal terms, for the dissipative thermodynamic sources we have
\begin{eqnarray}
\vec{R}^{M}&=&-\frac{\chi}{\tau_{M}}(\vec{B}-\vec{B}_{e}), \label{RM1}\\
\vec{R}^{J}_{i}&=&-\frac{\rho_M}{\tau_{J}}\vec{v}_{i}. \label{RJi}
\end{eqnarray}
Here $\tau_{M}$ and $\tau_{J}$ are new relaxation times associated with $\vec{M}$, and $\vec{J}$.  To ensure non-negative entropy production these times are non-negative.

The right-hand-sides of eqs.~(\ref{Jsi}-\ref{RJi}) are zero in equilibrium, consistent with no flux or source terms in equilibrium.

\subsection{Magnetic Equations of Motion}
\label{ss:mageqsmot}
From \eqref{Pij-J} and \eqref{rho-alpha}, we have $\vec{\Pi}_{ji}=\vec{\Pi}^{'}_{ji} -\alpha(\vec{B}-\vec{B}_{e})\delta_{ij}  =\vec{\Pi}^{'}_{ji} -(-\frac{\gamma\hbar}{2m}\rho_{M})(\vec{B}-\vec{B}_{e})\delta_{ij}$.  On dropping the dissipative term $\vec{\Pi}^{'}_{ji} $, we have $\vec{\Pi}_{ji} \approx\delta_{ij}\frac{\gamma\hbar}{2m}\rho_{M}(\vec{B}-\vec{B}_{e})$. We now consider the case of uniform $\vec{B}_{e}$, so $\partial_{i}\vec{B}_{e}=0$.  Then we take
\begin{equation}
\partial_{i}\vec{\Pi}_{ij}=\frac{\gamma\hbar}{2m}\rho_{M}\partial_{i}\vec{B}=\frac{\gamma\hbar}{2m}\rho_{M}\frac{1}{\chi}\partial_{i}\delta\vec{M}
=G\partial_{i}\delta\vec{M},
\qquad G\equiv\frac{\gamma\hbar}{2m}\frac{\rho_{M}}{\chi}.
\label{partialPi}
\end{equation}
Then, with $\partial_{t}\vec{M}=\partial_{t}\delta\vec{M}$, the equations of motion \eqref{dMdt} and \eqref{dJMdt} read
\begin{equation}
\partial_{t}\delta\vec{M}+\partial_{i}\vec{J}_{i}=-\gamma\delta\vec{M}\times\vec{B}-\frac{1}{\tau_{M}}\delta\vec{M}.
\label{dM/dt-time}
\end{equation}
\begin{equation}
\partial_{t}\vec{J}_{i}+G\partial_{i}\delta\vec{M}=-\gamma\vec{J}_{i}\times(\vec{B}+\lambda\vec{M})
-\frac{1}{\tau_{J}}\vec{J}_{i}.
\label{dJM/dt-time}
\end{equation}
In Eq. \eqref{dJM/dt-time}, we retain the dissipative part $\vec{J}^{\,'}_{i}$.

Clearly the independent variables are $\delta\vec{M}$ and $\vec{J}_{i}$.
Relative to the Fermi liquid theory of Ref.~\cite{Leggett70}, Eq.~\eqref{dM/dt-time} has $\delta\vec{M}$ in place of $\vec{M}$, and the decay term in $\tau_{M}$, and   Eq.~\eqref{dJM/dt-time} has $\partial_{i}\delta\vec{M}$ in place of $\partial_{i}\vec{M}$.

We now consider a time-dependence $e^{-i\omega t}$.  Then taking the time derivatives gives a factor of $-i\omega$, and we obtain
\begin{eqnarray}
\partial_{i}\vec{J}_{i}&=&-\delta\vec{M}\times\vec{B}-\left(\frac{1}{\tau_{M}}-i\omega\right)\delta\vec{M}.
\label{dM/dt-time2}\\
G\partial_{i}\delta\vec{M}&=&-\gamma\vec{J}_{i}\times(\vec{B}+\lambda\vec{M})
-\left(\frac{1}{\tau_{J}}-i\omega\right)\vec{J}_{i}. \quad
\label{dJM/dt-time2}
\end{eqnarray}

\subsection{Paramagnet}
\label{ss:Paramagnet}
Paramagnets in a magnetic field have the same magnetic symmetry as a ferromagnet, so the present theory should be applicable to paramagnets.  This invites comparison of the present theory to that of Leggett.\cite{Leggett70}

We write $\chi$ in terms of $N(0)$, the total density of states at the Fermi level:\cite{LandauFLT56,AbrikosovKhalatnikov58,BaymPethick78}
\begin{equation}N(0)=\frac{m^{*}p_{f}}{\pi^{2}\hbar^{3}},
\label{N0}
\end{equation}
\begin{equation}
\chi=\left(\frac{\gamma\hbar}{2}\right)^{2}\frac{N(0)}{1+\frac{1}{4}Z_{0}}.
\label{chiFL}
\end{equation}
For thermodynamic stability ($\chi\ge0$) we must have $1+Z_{0}/4\ge 0$, as for the alkali metals. Use of Leggett's results\cite{Leggett70} then gives
\begin{eqnarray}
G&=& \frac{1}{3}v_{F}^{2}\left( 1+\frac{1}{4}Z_{0} \right)\left( 1+\frac{1}{12}Z_{1} \right),
\label{rhoM}\\
\lambda&=&\frac{1}{\gamma \hbar}\frac{1}{N(0)}\left(Z_{0}-\frac{1}{3}Z_{1}\right),
\label{lambdaFL}
\end{eqnarray}
where $v_F$ is the Fermi velocity, and $Z_0$ and $Z_1$ are the first two coefficients of the Legendre expansion of the Fermi liquid interaction.  In the absence of such interactions, $\chi$ is the non-interacting value, $G=v_{F}^{2}/3$, and $\lambda=0$.

We then deduce that
\begin{equation}
\rho_{M}=\chi G=\left(\frac{\gamma\hbar}{2}\right)^{2}\frac{1}{3}v_{F}^{2}N(0)\left(1+\frac{1}{12}Z_{1}\right).
\label{rho_MFL}
\end{equation}

\section{Steady State Solutions}
\label{s:statics}
Among other things, spintronics requires the steady-state response of a ferromagnet when a spin-polarized current enters or leaves it.  In practice, for a thin film (our concern) the demagnetization field $\vec{B}_{d}$ will force $\vec{M}$ into the plane of the film, and the demagnetization field will be present and different for the two transverse directions.\cite{Kittel48}  However, we will neglect this asymmetry.

Recall that $\vec{M}=\vec{M}_{0}+\delta\vec{M}$, so $\delta\vec{M}$ and $\vec{J}_{i}$ are
determined by their respective boundary conditions.  Note that $G\equiv\rho_{M}/\chi$, which is given for the paramagnet by \eqref{rhoM}, can be thought of as the spin analog of the compressibility $\partial P/\partial\rho$. Without distinguishing between the longitudinal ($L$) and transverse ($T$) cases, the time-independent equations of motion have the form
\begin{equation}
\partial_{i}\vec{J}_{i}=-\gamma\delta\vec{M}\times\vec{B}-\frac{1}{\tau_{M}}\delta\vec{M}.
\label{dM/dt1}
\end{equation}
\begin{equation}
G\partial_{i}\delta\vec{M}=-\gamma\vec{J}_{i}\times(\vec{B}+\lambda\vec{M}_{0})-\frac{1}{\tau_{J}}\vec{J}_{i}.
\quad G\equiv\frac{\rho_{M}}{\chi},
\label{dJM/dt1}
\end{equation}

We now ``tensorize'' in spin space (using boldface for tensors), with $L$ ($T$) for the longitudinal (transverse) components:
\begin{equation}
G\rightarrow \mathbf{G}\equiv G_{L}\hat{M}\hat{M}+G_{T}(\mathbf{1}-\hat{M}\hat{M}).
\label{Gtens}
\end{equation}
We adopt this form of tensorization because the responses in the directions longitudinal and transverse to $\hat M$ are inequivalent, but the responses in the two transverse directions are equivalent, and related by a rotation about $\hat M$.   In principle, $\rho_{M}$ also can be tensorized.) 
To be specific we take $\vec{B}$ and the equilibrium $\vec{M}$ to be along $\hat{z}$ and to be nearly uniform.

For the longitudinal response we have
\begin{eqnarray}
\partial_{i}{J}_{Li}&=&-\frac{1}{\tau_{ML}}\delta{M},
\label{dM/dtL}\\
G_{L}\partial_{i}\delta{M}&=&-\frac{1}{\tau_{JL}}{J}_{Li}.
\label{dJM/dtL}
\end{eqnarray}
Taking $\partial_{i}$ on \eqref{dJM/dtL} and substituting \eqref{dM/dtL} then gives
\begin{equation}
G_{L}\nabla^{2}\delta{M}=\frac{1}{\tau_{JL}\tau_{ML}}\delta {M}.
\label{dM/dtL2}
\end{equation}
Setting $\delta M, {J}_{Mi}\sim e^{i\vec{k}\cdot\vec{r}}$, we find that
\begin{equation}
k^{2}=-\frac{1}{G_{L}\tau_{JL}\tau_{ML}}\equiv k^{2}_{L}.
\label{kL2}
\end{equation}
The sign of the square root is chosen so that, if a steady-state spin current is injected on one side of the sample, then it will be exponentially attenuated at the other side.

We now turn to the static transverse response. Following Refs.~\cite{ZhangLevyFert02} and \cite{TaniguchiAPEx08}, for conducting ferromagnets, we  neglect $\vec{B}$, so the exchange field dominates.
Taking $\vec{M}$ to be nearly constant, and performing $\partial_{i}$ on \eqref{dM/dt1} while substituting \eqref{dJM/dt1}, gives an equation in $\vec{M}_{T}$ alone:
\begin{equation}
G_{T}\nabla^{2}\vec{M}_{T}=\frac{1}{\tau_{MT}}\left(\gamma\lambda\vec{M}_{T}\times\vec{M}_{0}+\frac{1}{\tau_{JT}}\vec{M}_{T}\right)
\label{Jeq}.
\end{equation}
Writing $\vec{M}_{T}=M_{x}\hat{x}+M_{y}\hat{y}\sim e^{i\vec{k}\cdot\vec{r}}$, and taking $\vec{M}$ for $\vec{M}_{0}$, and along $\hat{z}$, gives the two vector components
\begin{eqnarray}
\left(G_{T}k^{2}+\frac{1}{\tau_{JT}\tau_{MT}}\right)M_{x}&=&-\frac{\gamma\lambda M}{\tau_{MT}}M_{y},
\label{vecx} \\
\left(G_{T}k^{2}+\frac{1}{\tau_{JT}\tau_{MT}}\right)M_{y}&=&\frac{\gamma\lambda M}{\tau_{MT}}M_{x}.
\label{vecy}
\end{eqnarray}

Cross-multiplication and factoring out $M_{x}M_{y}$ leads to the transverse wavevectors
\begin{equation}
k^{2}=-\frac{1}{G_{T}\tau_{JT}\tau_{MT}}\pm i \frac{\gamma\lambda M}{G_{T}\tau_{MT}}\equiv k^{2}_{T}.
\label{kT2}
\end{equation}
In evaluating the actual $k$-vector, the sign of $\pm$ depends on the situation and the sign of $\lambda$.  For propagation along $+\hat{x}$, we have ${\rm Re} (k_{x})>0$ and ${\rm Im} (k_{x})>0$.

This result for $k^{2}_{T}$ has the same structure as Eq.~\eqref{eq:ZLF} found by Zhang, Levy, and Fert.\cite{ZhangLevyFert02,comment}
Thus the present approach, based on the macroscopic properties of the magnetization $\vec{M}$ and its current $\vec{J}_{i}$, has succeeded in reproducing a result that has been
effective in analyzing numerous spintronics experiments.\cite{TaniguchiAPEx08}

Relative to Ref.~\cite{ZhangLevyFert02}, for transverse ($T$) spin current decay, comparing \eqref{eq:ZLF} and \eqref{kT2} gives
\begin{equation}
\tau_{MT}\leftrightarrow\tau_{sf}, \quad G_{T}\tau_{JT}\leftrightarrow2D_{0}, \quad \gamma\lambda\leftrightarrow\frac{J}{\hbar}
\label{equiv}
\end{equation}
where $D_{0}$ is a spin diffusion constant.  Thus the effective diffusion constant is a combination of $G_{L}$, associated with energy of nonuniformity (e.g., compressibility), and $\tau_{JL}$, associated with decay of $\vec{J}_{i}$.

When a $\vec{B}$ parallel to $\vec{M}$ is included,
\begin{equation}
k^{2}=k_{T}^2\equiv -\frac{1}{G_{T}\tau_{JT}\tau_{MT}}\pm i \frac{1}{G_{T}}  \Big[\frac{1}{\tau_{MT}}(\gamma B+\gamma\lambda M)+\frac{1}{\tau_{JT}} \gamma B\Big].
\label{kT2B}
\end{equation}

\section{Finite Frequency Modes}
\label{s:finitefrequencyLT}
We now consider finite frequency modes by taking
$\delta\vec{M}, \vec{J}_{i}\sim e^{i(\vec{k}\cdot\vec{r} -\omega t)}$. Before tensorizing the parameters, the time-independent equations of motion are given by \eqref{dM/dt-time} and \eqref{dJM/dt-time}.  To obtain finite frequency results, consider the time-dependence $e^{-i\omega t}$, so taking the time derivatives gives a factor of $-i\omega$.

Therefore the finite frequency modes can be obtained by making the following replacements to the decay rates in the steady state equations and solutions:
\begin{align}
\label{}
 &\frac{1}{\tau_{M}} \rightarrow   \frac{1}{\tau_{M}}-i\omega ,\\
 &\frac{1}{\tau_{\vec{J}}}  \rightarrow    \frac{1}{\tau_{\vec{J}}}-i\omega.
\label{finitefreqcorr}
\end{align}

For the longitudinal response, performing these replacements in \eqref{kL2} gives the longitudinal wavevectors at finite frequency:
\begin{equation}
k^{2}=k^{2}_{L}\equiv -\frac{1}{G_{L}}\left(\frac{1}{\tau_{JL}}-i\omega\right)\left(\frac{1}{\tau_{ML}}-i\omega\right).
\label{kL2-time}
\end{equation}\\
At high frequency we find $\omega=\pm \sqrt{G_{L}}k$, corresponding to longitudinal magnetic sound.

For the transverse response, performing these replacements in \eqref{kT2B} gives the transverse wavevectors at finite frequency:
\begin{eqnarray}
k^{2}=k_{T}^2&=& -\frac{1}{G_T}(\frac{1}{\tau_{JT}}-i\omega)(\frac{1}{\tau_{MT}}-i\omega)\nonumber\\
&&\pm \frac{i\gamma}{G_{T}}  \Big[(\frac{1}{\tau_{MT}}-i\omega)(B+\lambda M)+(\frac{1}{\tau_{JT}}-i\omega) B\Big].\qquad
\label{kT2B-time}
\end{eqnarray}
At high frequency we find $\omega=\pm \sqrt{G_{T}}k$, corresponding to transverse magnetic sound.

If $\omega_{L}\equiv \gamma B$ and $\omega_{E}=\gamma\lambda M$ satisfy $\omega_{E}\gg \omega_{L}, \omega, \tau_{MT}^{-1}, \tau^{-1}_{JT}$, then by appropriate sign choice
\begin{equation}
\omega= \frac{G_{T}}{\omega_{E}}k^{2}.
\label{wk2}
\end{equation}
This regime of quadratic behavior is as expected.

At frequencies that exceed the inverse of the local equilibrium decay time $\tau_{\,le}^{-1}$, one is in the zero sound regime, to which the present theory does not apply.

\section{Longitudinal Spin Waves for Paramagnets}
\label{s:FLcoefs}
The present theory, for ferromagnets, also applies to paramagnets, where the Fermi liquid interaction coefficients are known.  We thus apply the above results to this case. In terms of dimensionless quantities expressed in terms of the total density of states $N(0)$, at least three dimensionless forms have been employed.

The Russian literature employs
\begin{equation}
F_{l}=N(0)f_{l}, \qquad Z_{l}=N(0)\zeta_{l}.
\label{RL}
\end{equation}
Leggett uses $Z_{l}$.

Baym and Pethick employ
\begin{equation}
F^{s}_{l}=F_{l}, \qquad F^{a}_{l}=\frac{Z_{l}}{4}.
\label{BP}
\end{equation}
These works emphasize $^{3}$He.

The velocity of longitudinal spin waves $v_{M}$ for small spin polarization, in the collision-dominated regime, is given by analogy to that for first sound.  We express $v_{M}$ both in terms of $Z_{n}$ and $F_{n}=Z_{n}/4$:
\begin{eqnarray}
v_M^2&=&G= \frac{1}{3}v_{F}^{2}\left( 1+\frac{1}{4}Z_{0} \right)\left( 1+\frac{1}{12}Z_{1} \right),\nonumber\\
 &=&\frac{1}{3}v_{F}^{2}\left( 1+F_{0}^a \right)\left( 1+\frac{1}{3}F_{1}^a \right).
 \label{v_M}
\end{eqnarray}

For liquid $^{3}$He, we take $m^{*}/m=1.43$ and $p_{F}/\hbar=k_{F}=0.76\times10^{10}$ m$^{-1}$,\cite{AbrikosovKhalatnikov58}
which give $v_{F}=112$ m/s.  Then,
taking values at pressure $P=0$ for $Z_{0}/4$ and $Z_{1}/4$,\cite{Corrucini71,BaymPethick78,Alvesalo81,Candela86}
we find that $v_{M}$ ranges from 21 m/s to 37 m/s. To include damping for liquid $^{3}$He, in \eqref{kL2-time} we let $\tau_{ML}\rightarrow\infty$.

Platzman and Wolff, who expanded on Silin to study transverse spin resonance in paramagnetic metals, employ\cite{PlatzmanWolff}
\begin{equation}
B_{l}=f^{a}_{l}N(0)\frac{2}{2l+1}=F^{a}_{l}\frac{2}{2l+1}=Z_{l}\frac{1}{2(2l+1)}.
\label{PW}
\end{equation}

Eq.~\eqref{v_M} also applies to simple metals.  To evaluate this for Na and K, we take $B_{0}$ and $B_{1}$ from Refs.~\cite{DuniferPinkelSchultz74} and \cite{PinkelSchultz78}. For Na, we take $v_{F}=1.07\times 10^{6}$ m/s and find $v_{M}=5.8\times10^{5}$ m/s.
For K, we take $v_{F}=0.86\times 10^{6}$ m/s, and find $v_{M}=4.3\times10^{5}$ m/s.  \\

We are aware of no experimental evidence for or against longitudinal spin waves in either ferromagnets or paramagnets.

\section{Summary and Conclusions}
\label{s:summary}
\vskip-0.2cm
Using the symmetry of a ferromagnet, which is independent of particle statistics, we have applied Onsager's irreversible thermodynamics to study the transverse and longitudinal response of ferromagnets.  Two relaxation times, a spin stiffness, and an exchange field enter the theory.  The theory applies for frequencies below those of the local equilibrium relaxation time $\tau_{le}$.

We have explicitly studied the dc response.  For flow transverse to the wavevector, the wavevector has the same structure as that of the two-magnetization theory of Zhang, Levy, and Fert, namely, a complex-valued square
of the transverse wavevector with real part from diffusion and imaginary part from precession.  That form has been applied, with success, in previous analyses of transverse spin current decay in ferromagnets.\cite{TaniguchiAPEx08}

The related theories of Ref.~\cite{PlatzmanWolff} and Ref.~\cite{WIlsonFredkin70} have been used to successfully analyze experiments on finite-frequency transverse spin waves in various paramagnetic metals.\cite{DuniferPinkelSchultz74,PinkelSchultz78}

In the limit of infinite $\tau_{M}$ the above theory applies to systems with the same symmetry as atomic gases, both bosons and fermions.\cite{MiyakeMullinStamp85,Levitov02,Laloe02,Clark02}  We believe the same is true of the Leggett theory\cite{Leggett70} after re-interpretation of the Fermi liquid terms.

\section*{Acknowledgements}
\label{s:acknow}
\vskip-0.2cm
C. S. was supported by the National Natural Science Foundation of China under Grant No. 12105094, and by the Fundamental Research Funds for the Central Universities from China.

\begin{appendix}

\section*{Appendix: On the Response Functions $\chi$ and $\rho_m$}
\label{appendix:A}


For small external field $\vec{B}_{e}$ we have, in a classical average (with subscript $0$ for quantities when $\vec{B}_{e}$ is neglected), a $\vec{B}_{e}$-induced magnetization deviation.
With ``{\it Tr}\,'' for the trace, and with all vectors along $\vec{M}_{0}$, we have
\begin{equation}
\langle \delta M\rangle\approx Z_{0}^{-1}
Tr(\delta Me^{-E_{0}/{k_{B}T}}e^{B_{e}M/{k_{B}T}})
\approx \frac{\langle \delta M (M_{0}+\delta M)\rangle_{0}}{k_{B}T} B_{e}
=\frac{\langle (\delta M)^{2}\rangle_{0}}{k_{B}T} B_{e}\equiv \chi B_{e}.
\label{chiB*}
\end{equation}
This indicates how to calculate $\chi$.
The details of $\langle (\delta M)^{2}\rangle_{0}$ depend on the unperturbed energy.

Also in \eqref{deps},  the term $\frac{2m}{\gamma\hbar}\vec{v}_{e,i}\cdot d\vec{J}_{i}$  corresponds in Boltzmann weighting $e^{-E/{k_{B}T}}$ to the fictitious energy term $-\frac{2m}{\gamma\hbar}\vec{v}_{e,i}\cdot \vec{J}_{i}$, where the negative sign ensures that in ``equilibrium'' $\vec{v}_{i}$ is ``along'' the bivector $\vec{J}_{i}$.  Here $\vec{v}_{e,i}$ may be thought of as an imaginary field (not unlike what is used in systems with spontaneous symmetry-breaking, such as superfluids and superconductors) that determines the generalized direction associated with the symmetry breaking.  In the present case the spin currents, unlike superfluid currents and superfluid charge currents, do have an infinite lifetime.  However, they last long enough to be measured and in steady-state they decay in space over a finite distance, as discussed in the Introduction.

By analogy to the calculation for the actual field $\vec{B}_{e}$, for the bi-vector ``spin velocity field'' $\vec{v}_{e,i}$, on neglecting the dissipative part $\vec{J}'_{i}$ we have
\begin{equation}
\langle J_{\alpha i}\rangle
\approx Z_{0}^{-1}
Tr(J_{\alpha i}e^{-E_{0}/{k_{B}T}}e^{\frac{2m}{\gamma\hbar}v_{\beta j}J_{\beta j}/{k_{B}T}})
\approx \frac{1}{9}\frac{2m}{\gamma\hbar}\frac{\langle |\vec{J}_{i}^{2}|\rangle_{0}}{k_{B}T}{ v_{e,\alpha i}}
  \equiv\rho_{m}(\frac{\gamma\hbar}{2m})v_{e,\alpha i}.
\label{chiJveci}
\end{equation}
The $1/9$ occurs because we average over space and spin coordinates that are taken to be independent.
This indicates how to calculate $\rho_{m}$.  Again, the details will depend on the unperturbed energy.

\end{appendix}

{}


\begin{thebibliography}{11}
\bibitem{twomagvariables} The two magnetic variables are the usual magnetization $\vec{M}$ and the out of equilibrium magnetization $\vec{m}$ known as the spin accumulation.  In this model the spin current $\vec{J}_{i}$ is proportional to the gradient of $\vec{m}$. In the model of [\onlinecite{ZhangLevyFert02}], $\vec m$ is due to the conduction electron states ($s$ states) and $\vec M$ is due to the more localized $d$ electron states. Here by the two-magnetization model we mean a model in a more general sense with physical correspondences of spin accumulation $\vec{m}$ and magnetization $\vec{M}$ not specified. 
\bibitem{LeggettMdecay} Although Leggett was well aware of the decay of $\vec{M}$, for the small nuclear moment of $^{3}$He that decay is very slow on the time-scale of the spin-echoes in which he was interested.  For that reason decay of $\vec{M}$ was ignored.  With ratio $r$ of electronic moments to nuclear moments the relative electron-to-nucleon magnetic field energy scale varies as $r$ but the decay rate for $\vec{M}$ varies as $r^{2}$.  Since $r$ is large, decay of $\vec{M}$ is relatively more important for the electronic case.
\bibitem{fictitious} Fictitious order-producing external fields previously have been introduced in discussing superflow of mass in superfluids and superflow of charge (supercurrents) in superconductors.  In these cases the flow does not decay; in the present case the spin currents do decay, but they are long-lived enough to be observed.
\bibitem{vecmu} On introducing the polarization density $\vec{n}$ by $\vec{M}=-(\gamma\hbar/m)\vec{n}$ and the spin chemical potential $\vec{\mu}$ by $-\vec{B}^{*}\cdot d\vec{M}=\vec{\mu}\cdot d\vec{n}$ we obtain $\vec{\mu}=(\gamma\hbar/m)\vec{B}^{*}$, 
which is an alternate way of viewing the variables.  Here $\vec{\mu}$, an energy, is distinct from the spin accumulation $\vec{m}$, a magnetization.
\bibitem{comment}  In Ref.~\cite{ZhangLevyFert02} the quantity corresponding to $k^{2}_{T}$ is $-l^{-2}$, where $l$ is the complex decay length introduced after their Eq.~(18).
\end{thebibliography}
\end{document}

\bibitem{Silin1a} V. P. Silin, Zh. Eksp. i Teor. Fiz. 33, 127 (1957), Soviet Phys. JETP 6, 945 (1958), ``Oscillations of a Fermi liquid in a magnetic field.'' See also V. P. Silin, Zh. Eksp. i Teor. Fiz. 33, 495 (1957), Soviet Phys. JETP 6, 387 (1958), ``Theory of a degenerate Fermi liquid.''

\bibitem{Silin1b} V. P. Silin, Zh. Eksp. i Teor. Fiz. 35, 1243 (1958), Soviet Phys. JETP 8, 870 (1958), ``Oscillations of a Degenerate Fermi liquid.''

\bibitem{Slonc96} J. C. Slonczewski, J. Magn. Magn. Mater. 159, L1 (1996), ``Current-driven excitation of magnetic multilayers.''

\bibitem{Saslow17} W. M. Saslow, Phys. Rev. B 95, 184407 (2017), ``Irreversible thermodynamics of uniform ferromagnets with spin accumulation: Bulk and interface dynamics.''

\bibitem{SunSas19} C. Sun and W. M. Saslow, Phys. Rev. B 99, 104435 (2019), ``Transverse surface modes in ferromagnets: Coupled $\vec{M}$ and $\vec{m}$.''  


\bibitem{DyakonovPerel1} M. I. Dyakonov and V. I. Perel, Sov. Phys. JETP Lett. 13, 467 (1971), ``Possibility of Orienting Electron Spins with Current.''

\bibitem{DyakonovPerel2} M. I. Dyakonov and V. I. Perel, Phys. Lett. A 35, 459 (1971),  ``Current-induced Spin Orientation of Electrons in Semiconductors.''

\bibitem{Dyakonov07} M. I. Dyakonov. Phys. Rev. Lett.  99, 126601 (2007),  ``Magnetoresistance due to Edge Spin Accumulation.''

\bibitem{Dyakonov09} M. B. Lifshits and M. I. Dyakonov. Phys. Rev. Lett. 103, 186601 (2009), ``Swapping Spin Currents: Interchanging Spin and Flow Directions.''


\bibitem{ALSH19} V. P. Amin, J. Li, M. D. Stiles, and P. M. Haney, Phys. Rev. B 99, 220405(R) (2019). ``Intrinsic spin currents in ferromagnets.''  

\bibitem{SasChenMs} W. M. Saslow and C. Sun, manuscript in preparation.


\bibitem{WilsonFredkin70} A. B. Wilson and D. R. Fredkin, Phys. Rev. B 2, 4556, (1970), ``Collective Oscillations in a Simple Metal. I. Spin Waves.''

\bibitem{DuniferPinkelSchultz74} G. L. Dunifer, D. Pinkel, and S. L. Schultz, Phys. Rev. B 10, 3159 (1974), ``Experimental determintion of the Landau Fermi-liquid-theory parameters: Spin waves in sodium and potassium.''  See p. 3177 for the Fermi liquid coefficient values for Na and K.

